\documentclass[apjl]{emulateapj}

\usepackage{amsmath}
\usepackage{graphicx}
\usepackage{epsfig,psfrag,epic,eepic}
\usepackage{textcomp}
\usepackage{times}

\slugcomment{Draft version, to be submitted to ApJ Letters}

\shorttitle{Extended Outer Regions of the Debris Ring Around HR~4796~A}
\shortauthors{Thalmann et al.}

\newcommand{\pannekoek}{1}
\newcommand{\toronto}{2}
\newcommand{\ethz}{3}
\newcommand{\princeton}{4}
\newcommand{\washington}{5}
\newcommand{\madrid}{6}
\newcommand{\subaru}{7}
\newcommand{\steward}{8}
\newcommand{\charleston}{9}
\newcommand{\goddard}{10}
\newcommand{\eureka}{11}
\newcommand{\mpia}{12}
\newcommand{\fizeau}{13}
\newcommand{\naoj}{14}
\newcommand{\ifahawaii}{15}
\newcommand{\tokyo}{16}
\newcommand{\jpl}{17}
\newcommand{\sinica}{18}
\newcommand{\physmath}{19}
\newcommand{\sapporo}{20}
\newcommand{\sendai}{21}

\newcommand{\noteone}{\ensuremath{^\textrm{(1)}}}
\newcommand{\notetwo}{\ensuremath{^\textrm{(2)}}}
\newcommand{\notethree}{\ensuremath{^\textrm{(3)}}}
\newcommand{\notefour}{\ensuremath{^\textrm{(4)}}}
\newcommand{\notefive}{\ensuremath{^\textrm{(5)}}}

\begin{document}

\title{Images of the Extended Outer Regions of the Debris Ring Around 
	HR~4796~A\altaffilmark{$\star$}}

\author{C. Thalmann\altaffilmark{\pannekoek},   % checked
	M. Janson\altaffilmark{\toronto},    % checked
	E. Buenzli\altaffilmark{\ethz},    % special for this paper
	T.~D. Brandt\altaffilmark{\princeton},   % checked
	J.~P. Wisniewski\altaffilmark{\washington}, % special for this paper
	A. Moro-Mart\'in\altaffilmark{\madrid},    % checked
	T. Usuda\altaffilmark{\subaru},   % checked
	G. Schneider\altaffilmark{\steward},   % special for this paper
	J. Carson\altaffilmark{\charleston},    % checked
	M.~W. McElwain\altaffilmark{\goddard},   % checked
	C.~A. Grady\altaffilmark{\eureka,\goddard}, % special for this paper
	M. Goto\altaffilmark{\mpia},    % checked
%---
	L. Abe\altaffilmark{\fizeau},   % checked
	W. Brandner\altaffilmark{\mpia},   % checked
	C. Dominik\altaffilmark{\pannekoek},    % special for this paper
	S. Egner\altaffilmark{\subaru},   % checked
	M. Feldt\altaffilmark{\mpia},   % checked
	T. Fukue\altaffilmark{\naoj}, % special for this paper
	T. Golota\altaffilmark{\subaru},   % checked
	O. Guyon\altaffilmark{\subaru},   % checked
	J. Hashimoto\altaffilmark{\naoj},   % checked
	Y. Hayano\altaffilmark{\subaru},   % checked
	M. Hayashi\altaffilmark{\subaru},    % checked
	S. Hayashi\altaffilmark{\subaru},    % checked
	T. Henning\altaffilmark{\mpia},     % checked
	K.~W. Hodapp\altaffilmark{\ifahawaii},   % checked
	M. Ishii\altaffilmark{\subaru},   % checked
	M. Iye\altaffilmark{\naoj},      % checked
	R. Kandori\altaffilmark{\naoj},    % checked
	G.~R. Knapp\altaffilmark{\princeton},   % checked
	T. Kudo\altaffilmark{\naoj},     % checked
	N. Kusakabe\altaffilmark{\naoj},   % checked
	M. Kuzuhara\altaffilmark{\naoj,\tokyo},    % checked
	T. Matsuo\altaffilmark{\naoj},    % checked
	S. Miyama\altaffilmark{\naoj},    % checked
	J.-I. Morino\altaffilmark{\naoj},   % checked
	T. Nishimura\altaffilmark{\subaru},   % checked
	T.-S. Pyo\altaffilmark{\subaru},   % checked
	E. Serabyn\altaffilmark{\jpl},    % checked
	H. Suto\altaffilmark{\naoj},    % checked
	R. Suzuki\altaffilmark{\naoj},    % checked
	Y.~H. Takahashi\altaffilmark{\tokyo},  % special for this paper
	M. Takami\altaffilmark{\sinica},   % checked (Michihiro)
	N. Takato\altaffilmark{\subaru},   % checked
	H. Terada\altaffilmark{\subaru},   % checked
	D. Tomono\altaffilmark{\subaru},   % checked
	E.~L. Turner\altaffilmark{\princeton,\physmath},   % checked
	M. Watanabe\altaffilmark{\sapporo},   % checked
%---
	T. Yamada\altaffilmark{\sendai},   % checked
	H. Takami\altaffilmark{\subaru},   % checked
	M. Tamura\altaffilmark{\naoj}   % checked
}

\altaffiltext{$\star$}{Based on data collected at Subaru Telescope, which
	is operated by the National Astronomical Observatory of Japan.}

\altaffiltext{\pannekoek}{Anton Pannekoek Astronomical Institute, 
	University of Amsterdam, Amsterdam, The Netherlands; \texttt{thalmann@uva.nl}.}
\altaffiltext{\toronto}{University of Toronto, Toronto, Canada.}
\altaffiltext{\ethz}{Institute for Astronomy, ETH Zurich, Zurich, Switzerland.}
\altaffiltext{\princeton}{Department of Astrophysical Sciences, Princeton University, Princeton, USA.}
\altaffiltext{\washington}{University of Washington, Seattle, Washington, USA.}
\altaffiltext{\madrid}{Department of Astrophysics, CAB - CSIC/INTA, Madrid, Spain.}
\altaffiltext{\subaru}{Subaru Telescope, Hilo, Hawai`i, USA.}
\altaffiltext{\steward}{Steward Observatory, The University of Arizona, Tucson AZ, USA.}
\altaffiltext{\charleston}{College of Charleston, Charleston, South Carolina, USA.}
\altaffiltext{\goddard}{Goddard Space Flight Center, Greenbelt, USA.}
\altaffiltext{\eureka}{Eureka Scientific.}
\altaffiltext{\mpia}{Max Planck Institute for Astronomy, Heidelberg, Germany.}
\altaffiltext{\fizeau}{Laboratoire Hippolyte Fizeau, Nice, France.}
\altaffiltext{\naoj}{National Astronomical Observatory of Japan, Tokyo, Japan}
\altaffiltext{\ifahawaii}{Institute for Astronomy, University of Hawai`i, Hilo, Hawai`i, USA.}
\altaffiltext{\tokyo}{University of Tokyo, Tokyo, Japan.}
\altaffiltext{\jpl}{Jet Propulsion Laboratory, California Institute of Technology, Pasadena CA, USA.}
%\altaffiltext{\osaka}{Osaka University, Osaka, Japan.}
\altaffiltext{\sinica}{Institute of Astronomy and Astrophysics, Academia Sinica, Taipei, Taiwan.}
\altaffiltext{\physmath}{Institute for the Physics and Mathematics of the Universe, University
	of Tokyo, Japan.}
\altaffiltext{\sapporo}{Department of Cosmosciences, Hokkaido University, Sapporo, Japan.}
\altaffiltext{\sendai}{Astronomical Institute, Tohoku University, Sendai, Japan}

\begin{abstract}\noindent
We present high-contrast images of HR~4796~A taken with Subaru/HiCIAO
in $H$-band, resolving the
debris disk in scattered light.  The application of specialized angular 
differential imaging methods (ADI) allows us to trace the inner edge of
the disk with high precision, and reveals a pair of
``streamers'' extending radially outwards from the ansae.  Using a simple 
disk model with a power-law surface brightness profile, we demonstrate 
that the observed streamers can be understood as part of the smoothly 
tapered
outer boundary of the debris disk, which is most visible at the ansae.  
Our observations are consistent with the expected result of
a narrow planetesimal ring being ground up 
in a collisional cascade, yielding dust with a wide range of grain sizes.  
Radiation forces leave large grains in the ring and push smaller 
grains onto elliptical, or even hyperbolic trajectories.
We measure and 
characterize the disk's surface brightness profile, and confirm the 
previously suspected offset of the disk's center from the star's
position along the ring's major axis.  Furthermore, we present first 
evidence for an offset along the minor axis.  Such offsets are commonly 
viewed as signposts for the 
presence of unseen planets within a disk's cavity.  Our images also 
offer new constraints on the presence of companions down to the planetary
mass regime ($\sim$9\,$M_\textrm{Jup}$ at 0\farcs5, 
$\sim$3\,$M_\textrm{Jup}$ at 1\arcsec).
\end{abstract}

%% Keywords should appear after the \end{abstract} command. The uncommented
%% example has been keyed in ApJ style. See the instructions to authors
%% for the journal to which you are submitting your paper to determine
%% what keyword punctuation is appropriate.


%% From the front matter, we move on to the body of the paper.
%% In the first two sections, notice the use of the natbib \citep
%% and \citet commands to identify citations.  The citations are
%% tied to the reference list via symbolic KEYs. The KEY corresponds
%% to the KEY in the \bibitem in the reference list below. We have
%% chosen the first three characters of the first author's name plus
%% the last two numeral of the year of publication as our KEY for
%% each reference.

%% Authors who wish to have the most important objects in their paper
%% linked in the electronic edition to a data center may do so by tagging
%% their objects with \objectname{} or \object{}.  Each macro takes the
%% object name as its required argument. The optional, square-bracket 
%% argument should be used in cases where the data center identification
%% differs from what is to be printed in the paper.  The text appearing 
%% in curly braces is what will appear in print in the published paper. 
%% If the object name is recognized by the data centers, it will be linked
%% in the electronic edition to the object data available at the data centers  
%%
%% Note that for sources with brackets in their names, e.g. [WEG2004] 14h-090,
%% the brackets must be escaped with backslashes when used in the first
%% square-bracket argument, for instance, \object[\[WEG2004\] 14h-090]{90}).
%%  Otherwise, LaTeX will issue an error. 

\section{Introduction}

\label{intro}

Vega-type debris disks were first identified by infrared (IR) excesses around nearby main sequence stars \citep{aum84}.  The dust content of these second-generation disk systems is believed to be continuously replenished via collisional breakup of remnant planetesimals (cf.\ \citealt{wya08}). 
%Some 15\% of nearby main sequence stars are believed to harbor debris disks.

Since the imaging of the $\beta$ Pic system \citep{smi84}, nearly two dozen nearby debris disks have been spatially resolved. The morphological appearance of resolved debris disks is predicted to be influenced by interactions between dust in the disk and nearby planets \citep{oze00,kuc03}, the local interstellar medium \citep{man09}, recent stellar flybys and binary companions \citep{wya99}, mutual grain collisions \citep{thebault08}, and interaction of dust with residual gas \citep{klahr05}. Many resolved systems exhibit all morphological structures predicted by these mechanisms \citep{sch99,kalas05,gol06,kal08}.
%, and possible proof of the planet-disk link has been observed in the Fomalhaut system 
%\citep{kal08}. 
The observable morphology of resolved, optically thin debris disks is also wavelength dependent, as different bandpasses sample different grain size distributions \citep{wya06}.

\begin{figure*}[t]
\centering
%\vspace*{1mm}
\includegraphics[width=\linewidth]{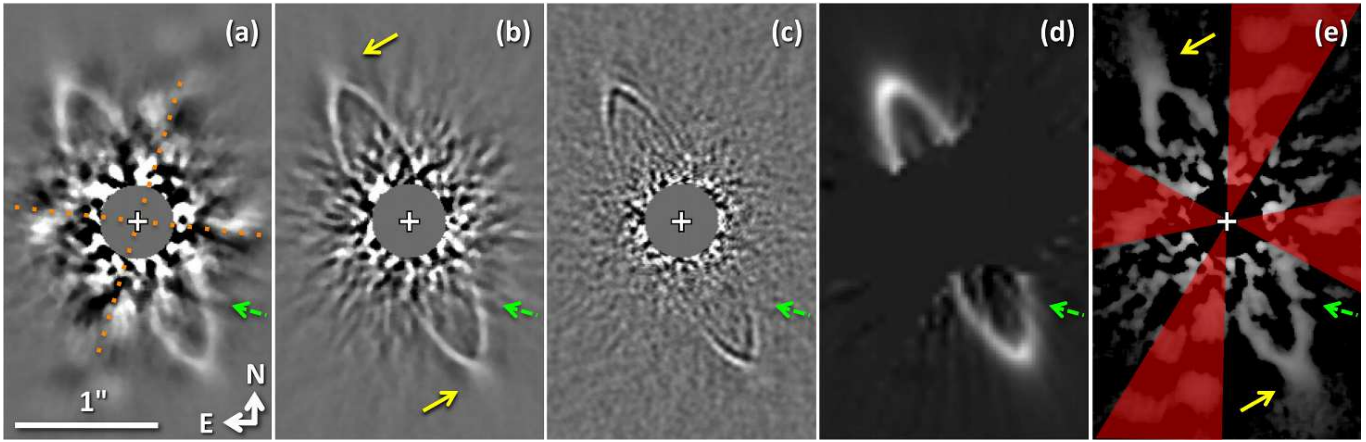}
%\vspace*{1mm}
\caption{High-contrast images of the HR~4796~A debris disk. White
	plus signs mark the star's location. The saturated central area is
	masked in grey.
	(a) Subaru HiCIAO $H$-band data after reduction with simple ADI
	(subtraction of a median background) and median
	smoothing on the scale of 5 pixels ($= 48$\,mas $\approx1\,\lambda/D$). Orange
	dotted lines mark the mean of the spider diffraction pattern.  The
	stretch is linear between $\pm3\!\cdot\!10^{-5}$
	times the stellar peak flux.
	(b) The same after ADI reduction with
	the LOCI algorithm, using conservative settings (optimization region
	area of 10,000 PSF footprints). Note the ``streamers'' extending radially 
	out from the ansae (yellow arrows).  Linear stretch of 
	$\pm1.5\cdot10^{-5}$.
	(c) The same with aggressive LOCI settings (300 PSF footprints)
	for point-source detection.
	Linear scale of $\pm6\cdot10^{-6}$.
	(d) HST/STIS 0.2--1.0\,$\mu$m 
	image from \citet[Figure 8]{schneider09} reprinted for
	comparison.	These authors point out a possible flux deficit in the 
	southwestern quadrant of the disk (dashed green arrows). 
	(e) Simple ADI image from (a) in logarithmic stretch 
	$[10^{-7},10^{-3}]$, showcasing the streamers. Red shading indicates 
	spider artifacts.
}
\label{f:images}
%\vspace*{2mm}
\end{figure*}

\object{HR 4796 A} is a young \citep[$\sim$8--10\,Myr;][]{stauffer95}, nearby \citep[72.8\,$\pm$\,1.7\,pc;][]{vanleeuwen07}, A0V-type star first identified as a debris disk system from an IR excess observed with IRAS \citep{jur91}. It has a co-moving M-type stellar companion at a separation of 7\farcs7 \citep{jur93}. The HR 4796 A debris disk has been spatially resolved at numerous optical, infrared, and sub-millimeter wavelengths \citep[e.g.,][]{koe98,jay98,sch99,sheret04,hinkley09,schneider09}, appearing as a narrow, highly inclined belt with a radius of 1\farcs05. Models suggest a dual dust population including gravitationally confined grains at $\sim$9--10\,AU in addition to those imaged at $\sim$70\,AU \citep{augereau99}, possibly involving radiationally evolved organic materials \citep{deb08}.  \Citet{wahhaj05} present multi-wavelength observations \looseness=-1
showing that the outer disk cannot be modeled as a single dust component; these authors adopt a two-component model with two different grain sizes.  Here we present new ground-based near-infrared imaging data at high spatial resolution, revealing the tapered outer regions of the debris disk.

 \section{Observations and data reduction}

Our observations of HR~4796~A were obtained with the Subaru
Telescope on May 24, 2011, within the SEEDS survey \citep[Strategic 
Exploration of Exoplanets and Disks with Subaru/HiCIAO,][]{tamura09}. The 
HiCIAO instrument \citep{hodapp08} with a 20\arcsec$\times$20\arcsec\ 
field of view and a plate scale of 9.50$\pm$0.02\,mas/pixel
was used. The image rotator operated in pupil-tracking mode to 
enable angular differential imaging \citep[ADI,][]{marois06}.  
A sequence of 260 images was taken in $H$-band with an 
exposure time of 10\,s, for a total integration time of 43.3\,min and a
total field rotation of 23\degr.  Weather conditions were good (seeing
0\farcs5--0\farcs8 in $V$-band), and the AO188 adaptive optics system 
\citep{minowa10} provided a FWHM of 6.5 pixels $=$ 62\,mas.

The images were corrected for flatfield and field distortion \citep{suzuki10}. 
Stellar position was estimated in each
frame by two different methods: (1) Fitting a Moffat profile to the PSF halo,
and (2) triangulating between symmetrical pairs of static speckles. Both 
were consistent with an empirical drift model 
($\vec{r}_i = \vec{r}_0 + \vec{v}_0 i + 1/2\,\vec{a_0} i^2$) plus
measurement noise. The difference 
between the two sets of centroids showed no systematic behavior, and was
consistent with incoherent combination of the two measurement noise 
sources.  We therefore used the drift model for image registration, for an
estimated centering accuracy of $\sim$0.3 pixels $=$ 2.9\,mas in the
co-added image.

ADI combined with the LOCI algorithm \citep[Locally Optimized Combination of 
Images,][]{lafreniere07} is currently the most successful ground-based 
imaging technique for the detection of planets 
\citep{marois10,lagrange10,currie11} and substellar companions \citep
{thalmann09,biller10,janson11}.  Furthermore, it has proven useful in revealing
high-contrast circumstellar disks \citep{thalmann10,buenzli10}.  

We applied three implementations of the
ADI technique to our data: (I) ``Simple ADI'', consisting of subtracting a 
median background from the entire dataset before derotating and co-adding.  
This method causes the least amount of flux loss and is therefore useful
for estimating the surface brightness profile of the disk. (II) 
``Aggressive LOCI'', using frame selection criteria of 0.5 FWHM (minimum
differential field rotation between images to be used
for mutual subtraction, to limit self-subtraction of physical
sources) and an
optimization region with an area of 300 PSF footprints.  Although most of 
the disk signal is lost and negative over-subtraction
effects appear, this method achieves excellent speckle 
suppression and provides the best constraints on point sources, such as 
planets. 
(III)  ``Conservative LOCI'', a compromise between the previous two methods, 
first described in \citet{thalmann10}.  This method preserves more disk
flux than aggressive LOCI while achieving significantly better speckle
subtraction than simple ADI.  The resulting image proves ideal for deriving
geometric parameters of the inner disk edge.  Due to the dataset's
limited field rotation budget, we do not use an increased 
frame selection criterion to conserve more disk flux, as \citet{thalmann10}
do.  Instead, we enlarge the
optimization area to 10,000 PSF footprints, lowering the impact of the disk
signal on the optimization process, as first demonstrated in 
\citet{buenzli10}.  Unless otherwise noted, we use the default numerical 
and geometric LOCI settings as defined in Table~1 and Fig.~1 of 
\citet{lafreniere07}.

\section{Results}

\subsection{Imaging the debris disk}

Figure~\ref{f:images} shows the results of applying the three ADI
reduction methods to our data: Simple ADI in (a),
conservative LOCI in (b), and aggressive LOCI in (c).  
In panel (d), we reprint the HST/STIS image by 
\citet[Figure 8]{schneider09} for comparison.  Starlight scattered from the
disk is visible as a strongly projected ellipse in all images, down to 
stellocentric separations of 0\farcs5 in simple 
ADI and 0\farcs4 in LOCI.  Differential imaging clearly 
resolves the inner edge of the disk.  

While most of the flux concentrates
around this edge, each ansa appears radially smeared out, extending a 
``streamer'' outwards along the major axis.  
Similar streamers have been observed in deep ADI or roll subtraction images 
of other
debris disks, such as HD~61005 \citep[``The Moth'',][]{buenzli10}
and Fomalhaut \citep{kalas10}.  They are thought to represent part of
the thin, extended, tapered outer reaches of the debris disk, which are 
most detectable at the ansae.  Oversubtraction makes them
appear narrower than they are.
In both simple ADI and LOCI, the flux at a given separation is unaffected
by any flux at smaller separations \citep{marois06,lafreniere07}, thus the
streamers cannot be artifacts caused by the bright ring further inwards.

\Citet{schneider09} note a marginal flux deficit in the southwestern 
quadrant of the debris disk.  Although there are some indications of
substructure in the same location in our dataset (dashed green
arrows),
the S/N ratio is insufficient to confirm the existence of a physical
disturbance in the disk.

%\subsection{Monte Carlo simulation for ring geometry fitting}
\subsection{Ring geometry from maximum regional merit fitting}

The crisp representation of the disk's inner edge in the conservative LOCI
image allows accurate measurement of its geometrical properties.  For that
purpose, we have devised the \emph{maximum regional merit} technique, 
which is loosely based on the method of \citet{buenzli10}.  

\begin{figure}[t]
\centering
%\vspace{2mm}
\includegraphics[width=\linewidth,trim=0mm 60mm 0mm 0mm]{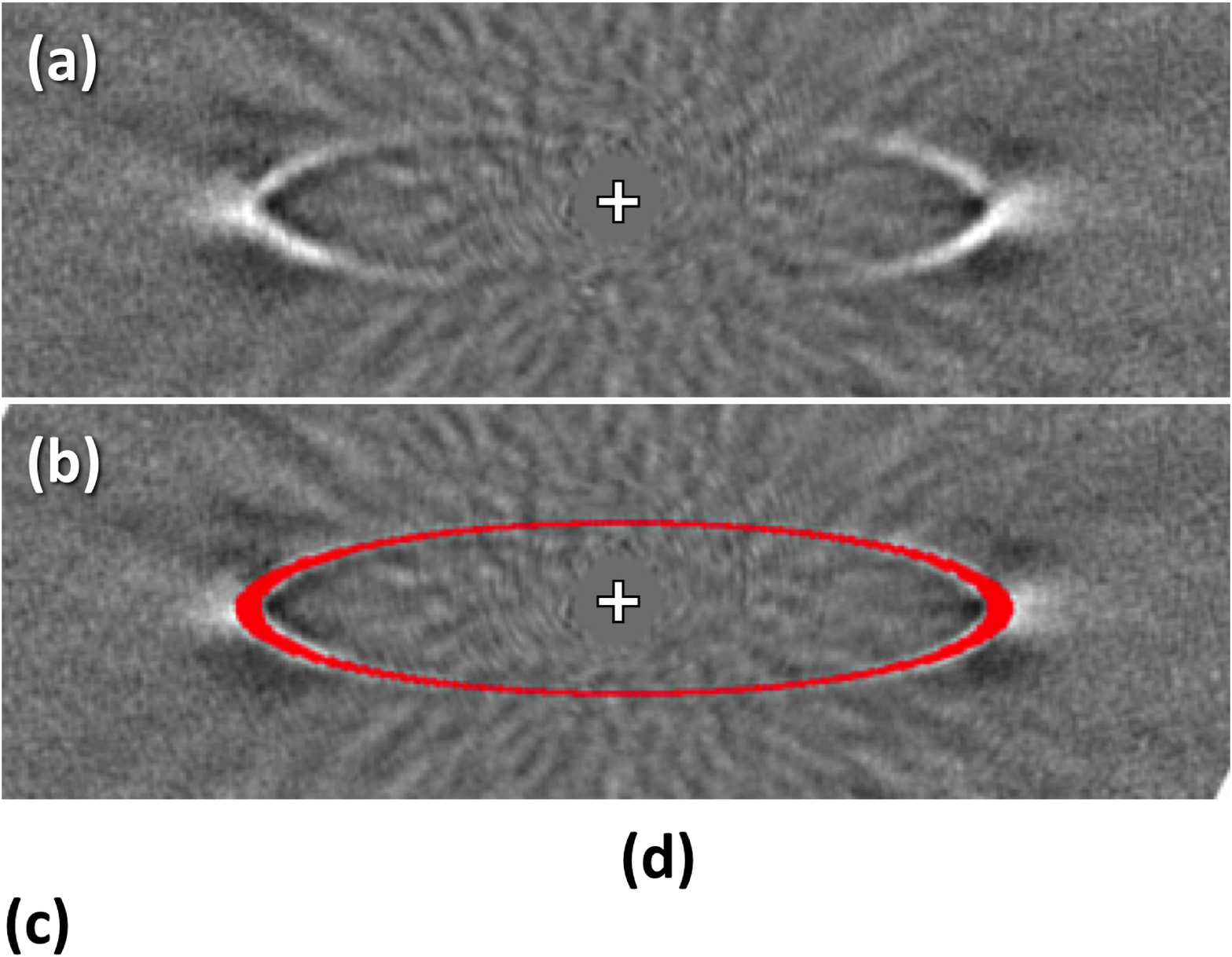}
\includegraphics[totalheight=0.57\linewidth, trim=5mm 5mm 0mm 0mm]{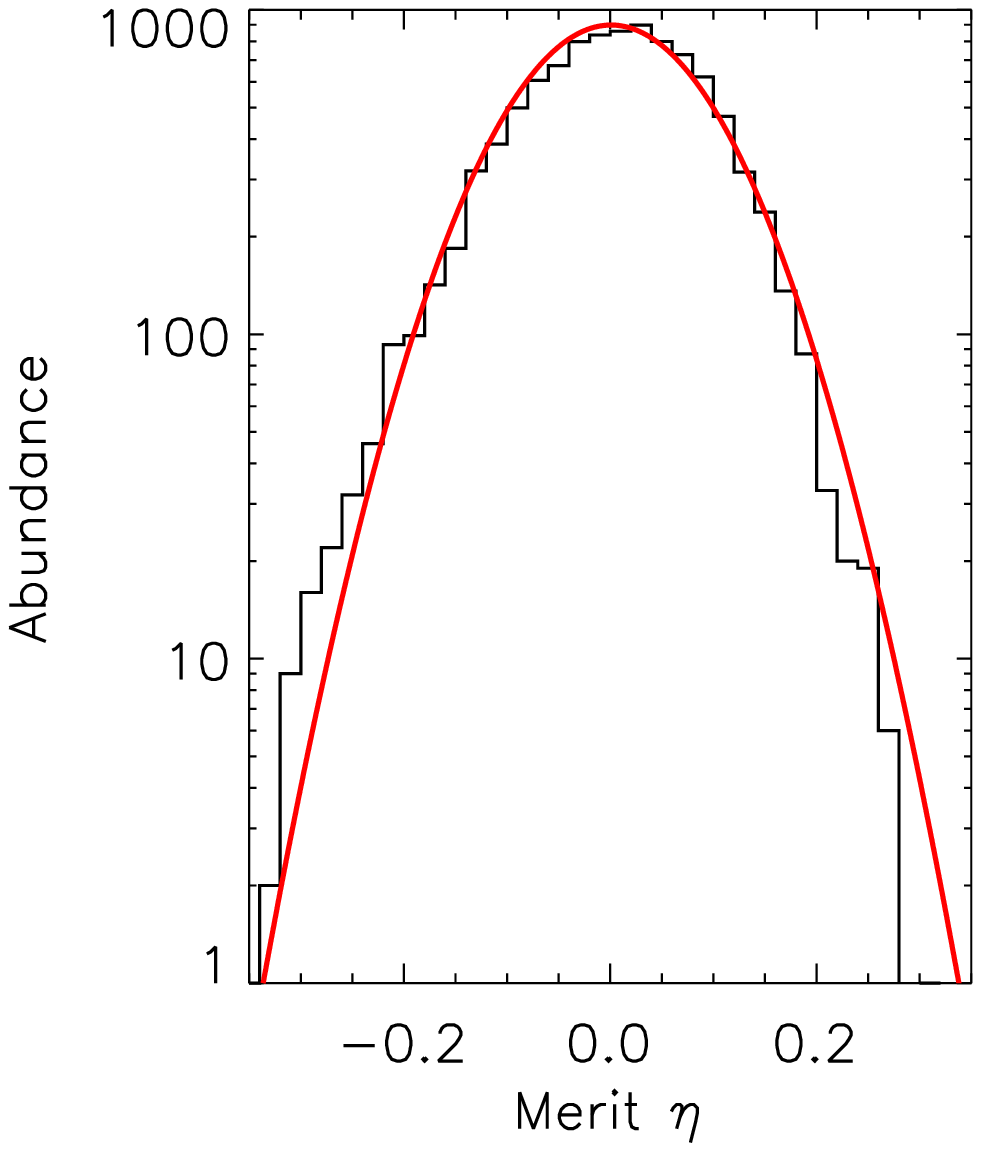}
\hspace{0.03\linewidth}
\includegraphics[totalheight=0.57\linewidth, trim=5mm 5mm 0mm 0mm]{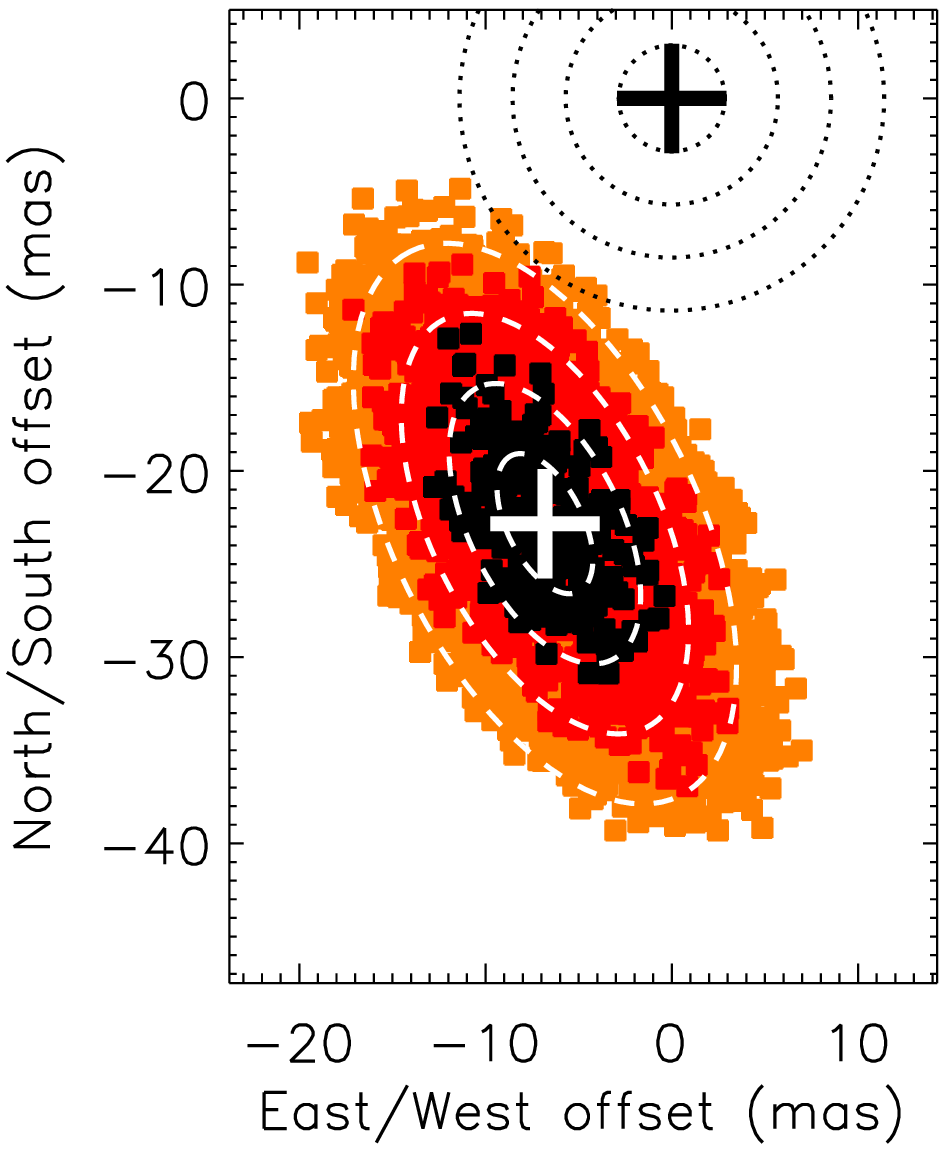}
%\vspace{1mm}
\caption{Derivation of geometric parameters. (a) Signal-to-noise map
of the conservative LOCI image of HR~4796~A, calculated in concentric
annuli around the star and excluding the bright ring from noise
estimation.  The stretch is $[-$10\,$\sigma$, 10\,$\sigma]$.  
(b) The same, but with red shading marking the sampling ellipse pixels
for the best-estimate ellipse.
%, which are defined as the mean values of the well-fitting solution family 
%(merits $\eta\ge\eta_\textrm{max}-\sigma_\eta$).  
(c) Histogram of merits $\eta$ measured in empty sky, showing a
quasi-Gaussian distribution.
(d) Scatter plot of the center offset parameters $(x,y)$ for the solution with 
merits $\eta$ within 1\,$\sigma_\eta$ (black), 2\,$\sigma_\eta$ (red), 
and 3\,$\sigma_\eta$ 
(orange) of the maximum merit value $\eta_\textrm{max}$, drawn from
Monte Carlo simulations of
300,000 parameter sets.  A white plus sign marks location of the best 
estimate $(\bar{x}, \bar{y})$; a black plus sign marks that of the star.  To
guide the eye, white dashed ellipses with semiaxes of 1--4 spatial standard
deviations of the 1\,$\sigma_\eta$ family along the major and minor axes
are provided, as well as dotted black circles marking the 1--4\,$\sigma$
uncertainty regions of the star's position.
%The spread of the solutions in $(x,y)$ parameter space is
%elliptical and scales linearly, showing that the errors can be treated as 
%Gaussian.
}
\label{f:fit}
\vspace*{2mm}
\end{figure}

First, we apply a median filter on the spatial scale of 5 pixels 
($\approx$1\,$\lambda/D$), which reduces the pixel-to-pixel noise while keeping 
larger edges and structures intact.  We then convert the image into a 
signal-to-noise map (S/N; Figure~\ref{f:fit}a) by calculating the standard 
deviation in 
concentric annuli around the star and dividing the pixel values in each
annulus by this noise level.  The disk itself is masked during
noise calculation.  The S/N map emphasizes the well-resolved outer parts of
the disk.

\begin{table}[t]
\vspace*{-3mm}
\caption{Numerical results on HR~4796~A}
\label{t:results}
\centering
%\begin{tabular*}{0.83\linewidth}{@{}@{\extracolsep{\fill}}l r@{.}l r@{.}l}  
\begin{tabular*}{\linewidth}{@{}l r@{ $\pm$ }l r@{ $\pm$ }l@{}}   %@{\extracolsep{\fill}}
\hline\hline
%Photometric position angle & 26.49\degr & 0.15\degr \\
Maximum merit ellipse size	
	& \multicolumn{2}{c}{ang.\ sep.} & \multicolumn{2}{c}{projected (AU)\noteone}\\
\hline
Semimajor axis $a$\notetwo   	& 1\farcs087 & 0\farcs023  & 79.2 & 1.7\\
Semiminor axis $b$  				& 0\farcs248 & 0\farcs007 & 18.0 & 0.5\\
\hline
Maximum merit ellipse orientation & \multicolumn{2}{c}{degrees\quad~}\\
\hline
Position angle $\Omega$			& 26.4 & 0.5 \\
Inclination $i=\arccos (b/a)$		& 76.7 & 0.5 \\
\hline
Maximum merit ellipse center
	& \multicolumn{2}{c}{ang.~sep.~(mas)} & \multicolumn{2}{c}{projected (AU)\noteone}\\
\hline
$x$ (along RA)				&$-$6.6 & 4.0\notethree & $-$0.48 & 0.21\notethree \\
$y$ (along Dec)				&$-$22.1 & 4.8\notethree & $-$1.61 & 0.28\notethree \\
$u$ (along major axis)\notefour		&$-$16.9 & 5.1 & $-$1.23 & 0.31 \\
$v$ (along minor axis)\notefour   	&15.8 & 3.6 & 1.15 & 0.16 \\
\hline
Upper limits on planets &
	\multicolumn{2}{c}{5\,$\sigma$ contrast} & 
	\multicolumn{2}{c}{Mass ($M_\textrm{Jup}$)\notefive} \\
\hline
At 0\farcs3 $=$ 18\,AU in proj.			& \multicolumn{2}{c}{5.8\,$\cdot$\,10$^{-4}$}
		& \multicolumn{2}{c}{$\sim$17}\\
At 0\farcs5 $=$ 36\,AU in proj. 					& \multicolumn{2}{c}{1.2\,$\cdot$\,10$^{-4}$}
		& \multicolumn{2}{c}{$\sim$9}\\
At 1\arcsec $=$ 73\,AU in proj.					& \multicolumn{2}{c}{5.7\,$\cdot$\,10$^{-6}$}
		& \multicolumn{2}{c}{$\sim$2.8}\\
At 2\arcsec $=$ 146\,AU in proj.				& \multicolumn{2}{c}{8.6\,$\cdot$\,10$^{-7}$}	
		& \multicolumn{2}{c}{$\sim$1.4}\\
\hline
\multicolumn{5}{@{}p{\linewidth}@{}}{\textsc{Notes.} 
(1) Conversion from arcsec to projected AU assuming a distance of 72.8\,pc; the
uncertainty of 1.7\,pc is not included in the errors.
(2) Based on the model disk described in Section~\ref{s:powerlaw}, the
photometric peak-to-peak radius $\hat{a}$ is $\sim$96\% of the 
%$a_\textrm{phot}$
maximum merit semimajor axis; this would yield a value of 
$\hat{a}=1\farcs044\pm0.023$ consistent with \citet{schneider09}.
(3) Errors in $(x,y)$
are correlated; use the uncorrelated errors in $(u,v)$ instead.
(4) The axes of the $(u,v)$ coordinate system are aligned with
the disk's major and minor axes, thus they are rotated counterclockwise
from the RA/Dec coordinate system $(x,y)$ by $\bar{\Omega}=26.4\degr$.
(5) Conversion from $H$-band contrast via the \texttt{COND}
evolutionary models \citep{allard01,baraffe03}, assuming an age of 
8\,Myr.}\\
\hline\hline
\end{tabular*}
\vspace*{2mm}
\end{table}

\begin{figure}[t]
\centering
%\vspace{2mm}
\includegraphics[width=\linewidth]{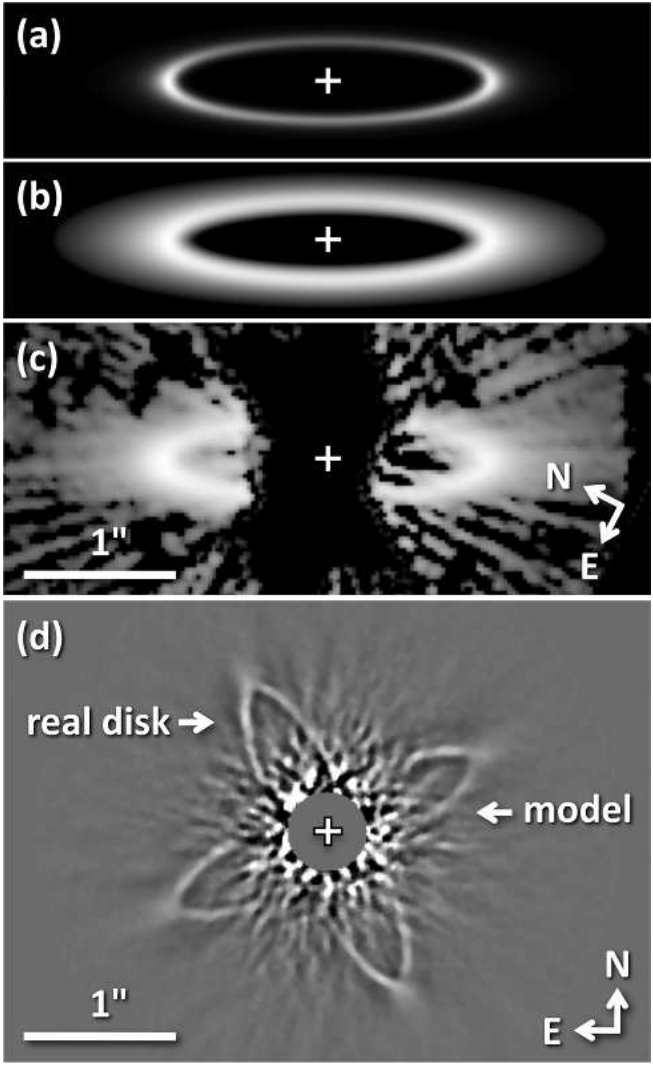}
%\vspace{2mm}
\caption{Modeling of the HR~4796~A debris disk. (a) Simulated
	scattered-light image of the simple power-law model disk described
	in the text, displayed in linear stretch.
	(b) The same in logarithmic stretch.  
	(c) PSF-subtracted 
	HST/STIS image from \citet{schneider09} in logarithmic stretch.  Note
	that a faint halo as predicted by the model is visible beyond the
	bright ring.
	(d) Result of conservative LOCI reduction of our Subaru HiCIAO data 
	after
	injection of the model disk at a 90\degr{} position angle offset
	(linear scale).
	The model appears qualitatively indistinguishable
	from the real disk, demonstrating that the morphology of the disk
	signal in our ADI images is well understood.
}
\label{f:model}
\vspace*{2mm}
\end{figure}

We then generate a large number of ellipses with randomized major and 
minor semi-axes $(a,b)$, position angles $(\Omega)$, and center positions
relative to the star $(x,y)$.  We assign a merit value $\eta$ to each set
of parameters, defined as the mean value of the pixels in the S/N map 
that lie within an ellipse-shaped sampling area described by the numerical
parameters.  The sampling area for a given parameter set $(a,b,\Omega,x,y)$
is defined as the area between two nested ellipses that share the same
orientation and center $(\Omega,x,y)$ but have semi-axes of 
$(a+\delta, b+\varepsilon)$ and $(a-\delta, b-\varepsilon)$, respectively.
We choose $\varepsilon=0.7$\,px, which results in a single-pixel thickness
along the minor axis, and $\delta=4$\,px, for a 7-pixel thickness along 
the major axis.  The value for $\delta$ corresponds to the observed width
of the high-flux areas in the ansae of the disk; this is to ensure that a
well-fitted sampling ellipse will be delimited by strong gradients in the
S/N image on all sides, and thus sensitive to displacement and scaling
(Fig.~\ref{f:fit}b).

We find that this method yields more accurate results for our data than the
conventional method of fitting of an ellipse to flux maxima in the ring 
\citep[as applied to HST data by][]{schneider09}, since our data has finer
spatial sampling but a higher photometric noise level than the HST data.
The disk profile is flat around the flux maxima and thus vulnerable to 
noise, while the tracking of sharp gradients is robust and profits from the
fine spatial sampling.

We find a maximum merit of $\eta_\textrm{max}=3.46$. In order to 
estimate the noise inherent to this merit figure, we measure $\eta$ for 
$10^4$ sampling ellipses placed in the empty sky region
90\degr{} away from the position angle of the disk.  We find a roughly 
Gaussian distribution (Fig.~\ref{f:fit}c) with a standard deviation 
$\sigma_\eta=0.077$ and a mean value $\mu_\eta=-0.0006$ ($\approx0$ 
within the errors).  We therefore
define the family of ``well-fitting'' solutions as those satisfying
$\eta\ge\eta_\textrm{max} - \sigma_\eta$, and the ``best estimate''
$(\bar{a}, \bar{b}, \bar{\Omega}, \bar{x},
\bar{y})$ as the mean value of the well-fitting solution 
family. 

Figure~\ref{f:fit}d plots the spread of the
center offset values $(x,y)$ of solutions within 1\,$\sigma_\eta$, 
2\,$\sigma_\eta$, and 3\,$\sigma_\eta$ of the maximum merit value 
$\eta_\textrm{max}$.  The contours
of those solution families are concentric ellipses with
axes aligned along the debris ring's axes, demonstrating 
approximately Gaussian error behavior.
To further test the validity of this maximum regional merit 
technique, we produced 10 model disks (as discussed in 
Section~\ref{s:powerlaw}) with random offsets from the star,
injected them into the empty sky region in our data, reduced them
with conservative LOCI, and retrieved the ellipse parameters as
described above. Deviations between measured and real offset
coordinates were consistently below $1\,\sigma$, with an RMS of
$0.4\,\sigma$.

The best-estimate parameters and their errors, including both
fitting errors and systematic errors from image registration and plate
scale, are presented in Table~\ref{t:results}.  
Most notably, we find that the ellipse center is offset along its major 
axis (towards the southwest) by $u=$ 16.9 $\pm$ 5.1\,mas, deviating from 
the star's position at 3.3\,$\sigma$
significance.  Since the major axis is unaffected by foreshortening
from projection along the line of sight, this translates directly 
into a physical distance of 1.2 $\pm$ 0.3\,AU, assuming an
approximately circular cavity.  
Combined with \citet{schneider09}'s independent measurement of 
19 $\pm$ 6\,mas (3.2\,$\sigma$ confidence), we can now confirm that 
HR~4796~A's debris disk is offset from its star.  Furthermore, our
data suggest an offset along the minor axis of 15.8 $\pm$ 3.6\,mas
in projection. If confirmed, this would correspond to a physical 
offset of $\sim$5\,AU.

\subsection{Constraints on planets}

Aggressive LOCI reduction of our image data does not detect any
planet candidates in the system, but imposes upper limits on the 
masses of 
potential unseen planets.  We use the photometric calibration 
method described e.g.\ in \citet{thalmann11}, 
estimating and correcting the separation-dependent partial flux 
loss from LOCI processing as described in \citet{lafreniere07}, and
converting the $H$-band flux into planet mass using the 
\texttt{COND} evolutionary models \citep{allard01, baraffe03}.
Table~\ref{t:results} lists the resulting 5\,$\sigma$ contrasts and 
planet masses for four separations.  The noise is calculated in 
concentric annuli, and the spider artifacts are masked out for this
purpose.

%\begin{figure}[t]
%\centering
%\vspace{2mm}
%\includegraphics[width=\linewidth]{f3.eps}
%\vspace{2mm}
%\caption{Constraints on point sources around HR~4796~A.  The red line
%marks the contrast of point sources detectable at 5\,$\sigma$ in the
%result of ADI reduction with aggressive LOCI.  }
%\label{f:contrast}
%\end{figure}

\section {Modeling} 
\label{s:modeling}

\subsection{Simple power-law model}
\label{s:powerlaw}

In order to understand the morphology of our disk images and gauge their usefulness for deriving disk properties, we generate a model disk by calculating synthetic scattered light images of inclined rings, with a power-law distribution of the scattering cross section $R$.
We assume the inclination of the disk to be given by the axis ratio, and construct a Keplerian disk with a small but nonzero eccentricity to match our observed semi-major axis offset (1.2\,AU, Table~\ref{t:results}). ÊThe main free parameters of the disk model are the radial power-law slopes, Ê
\begin{equation}
R(r) =  \left\{ \left( \frac{r}{a_0} \right)^{\displaystyle -2\alpha_\textrm{in}}  + \left( \frac{r}{a_0} \right)^{\displaystyle -2\alpha_\textrm{out}} \right\}^{\textstyle -0.5},
\end{equation}

\noindent
where $r$ is the radial distance to the star projected onto an ellipse with the desired geometry, $a_0$ is the semi-major axis, $\alpha_\textrm{in} > 0$ and $\alpha_\textrm{out} < 0$, and $R(r)$ is related to the surface density by $\sigma(r) = \sigma_0 \cdot R(r) \cdot \left(r/a_0\right)$.  We use a Henyey-Greenstein phase function and assume it is the same throughout the disk.  To compute images, we use the single-scattering radiative transfer code GRaTeR \citep{augereau99} to compute synthetic images. ÊThis approach is valid for an optically thin disk.

Given the scope of this model, we do not attempt to fit but merely approximate the real debris disk.  Our working parameters are $\alpha_\textrm{in} = 35$, indicating a very steep inner edge, $\alpha_\textrm{out}=-10$, and an asymmetry factor $g = 0.10$, implying a slight preference for forward scattering (cf.\ $g=0.16\pm0.06$ in \citealt{schneider09};
$g=$ 0.03--0.06 in \citealt{deb08}).

Figures~\ref{f:model}a and \ref{f:model}b show the simulated 
scattered-light images of the model disk at linear and logarithmic
intensity scale, respectively.  In Figure~\ref{f:model}c, we show 
the PSF-subtracted
image of the HST/STIS data from \citet{schneider09} in logarithmic
scale for comparison, revealing an extended halo beyond the bright
ring that matches the expected morphology.  Since this dataset is
reduced with PSF subtraction rather than ADI, it suffers from 
virtually no flux loss, but the background is dominated by residual
speckle noise.

To investigate the influence of ADI on disk flux, we inject the
simulated scattered-light images of the model disk into our raw
Subaru/HiCIAO dataset at a position angle offset by 90\degr{} with
respect to the real disk.  Figures~\ref{f:model}d
shows the results of conservative LOCI applied to this modified 
dataset.  The two disks appear virtually indistinguishable, 
demonstrating that our simple model adequately explains the observed 
morphology.

\begin{figure}[t]
\centering
%\vspace{2mm}
\includegraphics[width=\linewidth, trim=0mm 3mm 0mm 3mm]{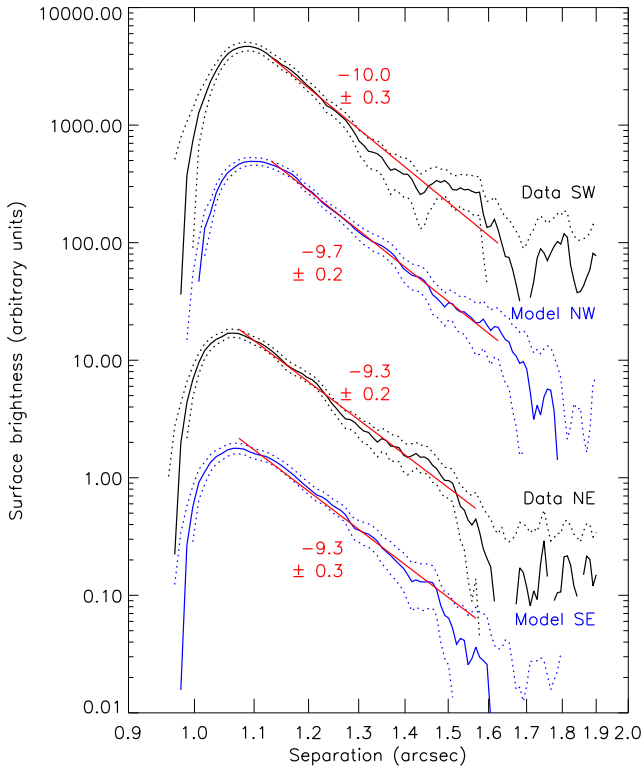}
%\vspace*{-3mm}
\caption{Surface brightness profiles of the HR~4976~A debris disk
	(black curves, dotted 1\,$\sigma$
	error bounds) extracted from 
	the ADI image in Figure~\ref{f:model} in strips 11 pixels wide
	oriented along the major axis. 
	The blue curves mark the corresponding profiles in the model 
	disk in the same image.  The latter demonstrate that the straight
	slope of the injected model disk is preserved in simple 
	ADI reduction.  
	%Thus, the apparent change of power-law slope around 
	%1\farcs3 observed on both sides of the real disk likely corresponds
	%to a physical feature.
	The red lines are linear fits to the plots in the interval
	1\farcs1--1\farcs6 with 
	respect to the disk center.  The numbers indicate their
	slopes (i.e., the power-law exponents), including fitting errors.
}
\label{f:slopes}
%\vspace*{10mm}
\end{figure}

\subsection{Surface brightness profiles}

We explore the viability of using the simple ADI image for the 
purpose of determining the $H$-band surface brightness profile of 
HR~4796~A's debris disk.  We extract a strip with
a width of 11 pixels
centered on the disk's major axis from the intensity image, and 
collapse it along the minor axis.  We repeat this for the simulated
disk image.  The results are shown in a log-log plot in 
Figure~\ref{f:slopes}.  The 1\,$\sigma$ error bounds are based on the
radial noise profile in empty sky quadrants of the simple ADI 
image before injecting the model disk.
Simple ADI preserves the model surface brightness profile.  The 
real disk's observed profile is consistent with a single power law.

\section{Discussion}

We have imaged and characterized the tapered outer boundary of 
HR~4796~A's debris disk using ground-based angular differential
imaging, and find that its $H$-band surface brightness profile 
is well described by a simple power law with a slope around
$-9.5$.  Our observations are consistent with the expected result
of a narrow planetesimal ring being ground up 
in a collisional cascade, yielding dust with a wide range of grain sizes.  
Radiation forces leave large grains in the ring and push smaller 
grains onto elliptical, or even hyperbolic trajectories.
Simulations of such this process 
predict a smoothly tapered outer boundary, consistent with our findings 
\citep[e.g.,][]{krivov06,strubbe06}. \Citet{wahhaj05} have demonstrated
that the disk cannot be modeled with a single dust grain size, 
implying that dust populations of varying grain size must be involved.
The brightness slope is a result of changing dust densities and grain 
properties with distance.  More detailed modeling is needed to
relate the observed slope to the disk's physical properties.
%Such efforts are underway and will be published in a future paper
%(van Lieshout et al.,\ in prep.).

%Observations of low-density disk components may provide important 
%constraints on dust blowout scenarios.  These components, most visible
%as ``streamers'' at the ansae of highly inclined disks, have 
%previously only been
%seen in a few cases \citep[e.g.,][]{buenzli10}. With the 
%improvements in high-contrast techniques and
%instruments that are presently underway, such features are likely to be
%more commonly detected in the future, which will enable the
%study of dust blowout and disk evolution across a broader sample. The ADI
%technique is generally well suited to detect these streamers.

Using our maximum regional merit technique,
we have corroborated the evidence for an offset between the
debris disk's inner edge and the star.
The dynamical influence of unseen planets orbiting within the disk 
cavity is most commonly invoked to explain such offsets \citep[e.g.,]
[and references therein]{kalas05, thalmann09, buenzli10}, and may be
the cause of the proposed but unconfirmed intra-ring gap.  
Our confirmation of the ring offset also speaks against a dynamically
cold source planetesimal population as proposed by \citet{thebault08}
as an alternative to the planet hypothesis for explaining the 
morphology of the ring.  The planets
may well be detectable with the upcoming next-generation high-contrast
imaging facilities.  
%\vspace*{-5mm}

%%%%%%%%%%%%%%%%%%%%%%%%%%%%%%%%%%%%%%%%%%%%%%%%%%%%%%%%%%%%%%%%%%%%%%%%
%%%%%%%%%%%%%%%%%%%%%%%%%%%%%%%%%%%%%%%%%%%%%%%%%%%%%%%%%%%%%%%%%%%%%%%%
%%%%%%%%%%%%%%%%%%%%%%%%%%%%%%%%%%%%%%%%%%%%%%%%%%%%%%%%%%%%%%%%%%%%%%%%
%%%%%%%%%%%%%%%%%%%%%%%%%%%%%%%%%%%%%%%%%%%%%%%%%%%%%%%%%%%%%%%%%%%%%%%%

\acknowledgements
We thank David Lafreni\`ere for generously providing us with the
source code for his LOCI algorithm, and Jean-Charles Augereau for his
GRaTer code.  The authors acknowledge partial support from the Swiss 
National Science Foundation (SNSF), US National Science Foundation 
grants AST-1009203 and DGE-0646086, and a Japanese MEXT Grant-in-Aid for
Specially Promoted Research (No.~22000005).

{\it Facilities:} \facility{Subaru (HiCIAO, AO188)}.

\clearpage

\end{document}